\documentclass[11pt,preprint]{aastex}
\newcommand{\be}{\begin{equation}}
\newcommand{\ee}{\end{equation}}
\newcommand{\bea}{\begin{eqnarray}}
\newcommand{\eea}{\end{eqnarray}}

\slugcomment{Submitted to ApJ ??}

\shorttitle{Damping of Wavws and Turbulence}
\shortauthors{Perosian, Lazarian \& Yan}
\begin{document}
\title{Damping of MHD turbulence in Solar Flares}
\author{Vah\'e Petrosian\altaffilmark{1}}
\affil{Center for Space Science and Astrophysics, Department of Physics, Stanford
University, Stanford, CA 94305}
\email{vahep@stanford.edu}
\author{ Huirong Yan\altaffilmark{2}and A. Lazarian}
\affil{Department of Astronomy, University of Wisconsin, Madison, WI}
\altaffiltext{1}{Department of Applied Physics, Stanford
University, Stanford, CA 94305}
\altaffiltext{2}{current address: CITA, University of Toronto, Canada }

\begin{abstract}

We describe the cascade of plasma waves or turbulence injected, 
presumably by reconnection, 
at scales comparable to the
size of a solar flare loop, $L\sim 10^9$ cm, to scales comparable to elementary particle gyro radii, 
and evaluate their damping  at small scales by various mechanisms. We 
show that the classical
viscous damping valid on scales larger than the collision mean free 
path ($\sim 10^8$ cm) is 
unimportant for magnetically dominated or low beta plasmas and the primary 
damping mechanism is the 
collisionless damping by the background particles. 
We show that the damping rate is proportional to the 
total random momentum density of the particles. For solar flare conditions 
this means that in 
most flares, except the very large ones, the damping is dominated by 
thermal background electrons.
For large flares one requires acceleration of essentially all 
background electrons into a nonthermal distribution
so that the accelerated electrons can be important in the damping of the 
waves. In general, damping by 
thermal protons is negligible compared to that of electrons except for 
quasi-perpendicular propagating 
waves. 
Damping due to nonthermal protons is also negligible compared to nonthermal electrons in
most flares which are electron dominated, except for  
rare proton dominated flares with strong nuclear gamma-ray line emission.

Thus for common 
flares collisionless damping by background thermal electrons is the 
primary damping mechanism.
Using the rate for this process we determine the critical scale (or wave vector $k_c$) below  which
({\it i.e.} for $k>k_c$) the damping becomes important and  the spectrum of the turbulence steepens.
This critical scale, however, has strong dependence on the angle of propagation of the waves with 
respect to the background magnetic field direction. The waves can cascade down to very small scales,
such as the gyro radii of the particles which are well beyond the MHD regime, 
at small angles (quasi-parallel
propagation) and possibly near 90 degree (quasi-perpendicular propagation). 
Thus, the spectral distribution would be highly 
anisotropic  at small scales.  

\end{abstract}

\keywords{plasmas --- turbulence --- acceleration of particles}

\section{INTRODUCTION}
The mechanism of energy  release and the process of its transfer to heating 
and
acceleration of nonthermal particles in many magnetized astrophysical 
plasmas in general, and solar 
flares in particular, are still matter of considerable debate. Recent 
research show that turbulence 
may play an essential role in these processes. In the case of solar flares, it is believed that the
energy comes  from release of stored magnetic energy via reconnection (see
Priest \& Forbes 2000, Lazarian, Vishniac \& Cho 2004 and discussions 
therein). Both the ordinary and magnetic 
Reynolds numbers $Re=l\delta v/\nu, Rm=l\delta v/\eta \gg 1$, so that the magnetized plasma develops 
turbulence. Here $\delta v$ is the velocity change across a turbulent region of scale 
$l$, and $\nu$ 
and $\eta$ are the  viscosity and magnetic diffusion coefficients, respectively. More importantly, recent
high resolution observations of solar flares by {\it Yohkoh} and {\it RHESSI} satellites have 
provided ample evidence that, at least from the point of view of particle acceleration, plasma
turbulence and plasma waves appear to be the most promising agent  not only for the
acceleration mechanism but also the general energizing of flare plasma (see {\it e.g.} 
Petrosian \& Liu 
2004, hereafter PL04, and references cited there). This may also be true in other 
situations (Liu, 
Petrosian \& Melia 2004). These investigations go beyond assuming, {\it e.g.} a power law 
electron distribution, as is commonly done, and calculate the expected spectrum based on
interaction of plasma particles with turbulence. However, in  such treatments the spectrum
$W(\mathbf k)$ of 
the turbulence as a function of wavevector $\mathbf k$ is an input parameter rather than 
evaluated based on first principles. The limitations of such an approach are self-evident.
Particle acceleration rate depends on the wave spectrum and the wave damping rate are
partially determined by the particle spectrum. In general, one requires
a self-consistent treatment of the coupled wave-particle kinetic equations describing
the generation of turbulence and its subsequent interactions with the 
background plasma. 

An attempt to solve the problem self-consistently was undertaken by Miller, LaRosa \& Moore (1996),
where coupled equations for energetic particles and turbulence have been studied. For our purposes
we can rewrite these equations in the following form:
\begin{eqnarray}
\label{kineq}
{\partial N \over \partial t}
& = & {\partial \over \partial E}\left[D_{EE}{\partial N \over \partial E}
 - (A-\dot E_L) N\right]
 - {N \over T^p_{\rm esc}} +{\dot Q}^p\nonumber \\
{\partial {W} \over \partial t}
& = & {\partial \over\partial k_i}\left[D_{ij}{\partial\over \partial k_j}{W}\right] - \Gamma({\mathbf k}){W} - {{W}\over T^{W}_{\rm esc}({\mathbf k})} + \dot{Q}^{W}.
\label{coupled}
\end{eqnarray}
Here $D_{EE}/E^2, A(E)/E$ and ${\dot E}_L/E$ give the diffusion, direct acceleration and energy loss 
rates of the particles, respectively, and 
$D_{ij}({\mathbf k})/k^2$ and  $\Gamma({\mathbf k})$ describe the 
cascade and damping rates of turbulence.
The $\dot{Q}$'s and the terms with the escape times $T_{\rm esc}$ describe the source and leakage of 
particles and waves. Note, that we use a more general form of the equation set compared to that in
Miller et al. (1996). In our treatment the anisotropy of the turbulent statistics is allowed and
the leakage of the particles as well as turbulent energy is accounted for. 
In general, and particularly in the case of solar flares that we deal here, it is reasonable to
assume  that the turbulence is generated mainly at a large scale 
$l$ comparable or somewhat 
smaller than the spatial extent $L$ of the region 
(with initial velocity  $\delta V$ and magnetic field 
fluctuation $\delta B$). The above equations then determine the resultant spectrum and other 
characteristics 
of the  turbulence, as it cascades to smaller scales and is damped by the background thermal plasma,
as well as the spectrum of the accelerated nonthermal particles.

In recent years there has been a substantial progress i) in the
understanding of cascade of incompressible (Goldreich \& Shridhar 1995) and
compressible MHD turbulence (see Cho \& Lazarian 2005a
and references therein), and damping of compressible MHD turbulence 
(Yan \& Lazarian 2004, henceforth YL04); and ii) in determination of plasma-wave-particle
and  MHD-turbulence-particle interaction rates
(see, {\it e.g.} Dung \& Petrosian 1994; Pryadko \& Petrosian 1997, 1998, 1999; PL04;
Chandran 2000; Yan \& Lazarian 2002, 2004, henceforth YL02, YL04, 
Cho \& Lazarian 2005b). These advances
allow a more thorough description of the coefficients involved in equation (\ref{kineq}).
In this paper we limit our attention to  one
aspect of this complex problem, namely, to damping of turbulence represented by 
the coefficients $\Gamma({\bf k})$.
After a brief review of recent progress in the understanding of the 
cascade process in \S2 we evaluate the damping rates due to thermal particles in the background 
plasma and a nonthermal population representing the accelerated spectrum $N(E)$ (\S3). 
In \S4 we summarize our results and discuss their use in our future
work on solving the coupled wave-particle kinetic equations.

\section{TURBULENCE AND ITS CASCADE}

Plasma turbulence can be decomposed into many wave modes with frequencies $\omega$ 
extending essentially from zero frequency to 
beyond the ion (in our case proton) and electron gyro frequencies 
$\Omega_p=eB/m_pc$ and  $\Omega_e=\Omega_p/\delta$  
($\delta=m_e/m_p$ is the electron to proton mass ratio), and wavevectors $k=1/l$ 
spanning the spatial scales from the injection scale to the gyro radius of electrons 
obeying a complex dispersion 
relation determined by the values of density $n$, temperature $T$ and magnetic field 
$B$ of the background plasma.

When we consider turbulence injected at a scale much larger
than the proton (or ion) skin depth $\sim v_A/\Omega_p=230(10^{10}{\rm cm}^{-3}/n)^{1/2}$ cm,
where 
\be\label{AlfvenV}
\beta_A=v_A/c= 7\times 10^{-3} (B/100 {\rm G})(10^{10}{\rm cm}^{-3}/n)^{1/2}
\ee
is the Alfv\'en velocity in unit of speed of light $c$,
initially we deal with modes for which the plasma acts as a single 
fluid and we are in the MHD regime where 
the dispersion relation simplifies considerably. It has been known for decades, 
that the weak MHD perturbations can be decomposed into
Alfv\'{e}nic, slow and fast waves with well-known dispersion relations (see {\it e.g.} Sturrock 
1994). However, it was also believed that such a decomposition does not make much
sense for a highly non-linear phenomenon of MHD turbulence, where the
modes were believed to be strongly coupled (see  Stone et al. 1999). 
A study of mode coupling in Cho \& Lazarian (2002 and 2003,
henceforth  CL02 and CL03) has shown that the coupling is appreciable only at the
injection scale, while along the cascade
to smaller scales the transfer of energy between the
modes is suppressed%
\footnote{An intuitive insight into this process can
be traced to Goldreich-Shridhar (1995) study (see
also Lithwick \& Goldreich 2001, CL02).}. 
This justifies the decomposition to different modes even for strong MHD turbulence (see CL02, CL03)
and allows us  to treat their cascade, and interactions with
charged particles, of Alfv\'enic, fast and slow modes  separately (see YL04, YL03). 

In solar flares, and in many other astrophysical plasmas, one is dealing with a magnetically 
dominated plasma with the plasma beta parameter
\be\label{beta}
\beta_p=8\pi nkT/B^2=3.4\times 10^{-2}(n/10^{10} {\rm cm}^{-3})(100 {\rm G}/B)^2(T/10^7{\rm K})\ll 1,
\ee
 which also means that the
Alfv\'{e}n speed  is greater than the sound speed. In this case the slow mode can be ignored (see Cho 
\& Lazarian 2005b)
and one can use the cold plasma dispersion relation 
\be\label{disp}
\omega=v_A k \cos{\theta} \,\,\,\,\,\,  {\rm and} \,\,\,\,\, \omega=v_A k 
\ee
for the Alfv\'{e}n and fast modes, respectively, where 
$\theta$ is the angle of propagation of the wave with respect to the magnetic field%
\footnote{In very highly magnetized plasma $\beta_A^2\gg 1$, $v_A$ in equation (\ref{disp})
should be replaced by 
$v_A/\sqrt{v_A^2+c^2}$, and the actual dispersion relations deviate from these at 
shorter scales or for $k>k_{crit}\sim \Omega_p/(c\beta_A^2)(1+\cos ^2 
{\theta}/\delta)^{1/2}$. Fr solar flare conditions this scale is larger than the proton gyroradius
$v_{th}/\Omega_p\sim 30$ cm so that the above dispersion relations are good approximations }. 

Turbulence generated at large scales can cascade to small scales by nonlinear interactions.
One important characteristic of turbulence is its self-similarity.
Power law spectra were obtained numerically for Alfv\'{e}nic, 
fast and slow
mode turbulence in CL02 and CL03 for the case when turbulent energy is
injected at large scales. It has also been demonstrated that
Alfv\'en (and slow) modes exhibit scale-dependent anisotropy similar to that described 
by Goldreich \& Sridhar (1995) for incompressible turbulence. 

This can be understood on a qualitative level as follows. For Alfv\'enic turbulence,
the mixing motions perpendicular to the magnetic field
couple with the wave-like motions parallel to the magnetic field providing
so-called critical balance condition, 
$k_\perp v_{k}\sim k_{\parallel}v_{A}$, where $k_{\parallel}$ and $k_\perp$ are the
parallel and perpendicular components of the total wave vector $\mathbf{k}$. This, when combined
with the Kolmogorov scaling for mixing motions 
with $v_{k}\simeq \delta V(k_\perp L)^{-1/3}$ (see Lazarian \& Vishniac 1999), yields 
a scale dependent anisotropy 
$k_{\parallel}L\sim (\delta V/v_A)(k_{\perp}L)^{2/3}$. The mixing motions associated with
Alfv\'enic turbulence induce the scale-dependent anisotropy on slow modes,
which on their own would evolve on substantially longer time scale.
The anisotropic spectrum of the Alfv\'enic turbulence can be described as
(Cho, Lazarian \& Vishniac 2003)%
\footnote{Note that integrating over the parallel and perpendicular components one gets 
$W(k_\perp)\propto k_\perp^{-5/3}$ and $W(k_\parallel)\propto k_\parallel^{-2}$, respectively.}
\be
\label{Alfspec}
W(k_{\parallel},k_{\perp}) = \frac{L^{-1/3}}{6\pi}k_{\perp}^{-10/3}
\exp(-L^{1/3}|k_{\parallel}|/k_{\perp}^{2/3}).
\ee

For Alfv\'en and slow modes the cascade time, same as the hydrodynamic eddy turnover time, is 
\be
\label{tcasAlf}
\tau_{cas}\simeq l/v_{k}\simeq \tau_0(k_{\perp}L)^{-2/3}/M_A\simeq 
\tau_0(k_{\parallel}L)^{-1},
\ee
where we have defined a characteristic time and the Alfv\'en Mach number at injection
\be\label{tau0} 
\tau_0=L/v_A=4.5(L/10^9{\rm cm})(100{\rm G}/B)(n/10^{10}{\rm cm}^{-3})^{1/2} {\rm s}\,\,\,\,
{\rm and}\,\,\,\, M_A\equiv\delta V/v_A.
\ee

Fast modes in low $\beta_p$ plasma, on the other hand, 
develop on their own, as their phase
velocity is only marginally affected by the mixing motions induced
by Alf\'ven modes.   According
to CL02, fast modes 
follow an isotropic ``acoustic'' cascade%
\footnote{This scaling,
plausible on theoretical grounds (CL03, Cho \& Lazarian 2005a), 
is supported by numerical simulations of CL03, but may still require
more testing. On the other hand, in the high
$\beta_p$ case the scaling of the fast modes trivially reduces to the scaling 
of acoustic turbulence because the
fast modes are essentially acoustic fluctuations.} with $W(k)\sim k^{-3/2}$.
 
For such a cascade in each wave-wave collision a small fraction 
of energy equal to 
$v_{ph}/v_k$ is transfered to smaller scales so that the 
cascade time scale is 
characterized by (CL02):
\be
\label{tcasfast}
\tau_{cas}=(v_{ph}/v_k)(l/v_{k})=\tau_0(kL)^{-1/2}/M^2_A.
\ee
Here $v_{ph}=\omega/k=v_A$ is the phase velocity of the fast 
mode, and  we have used the scaling relation $v_k=\delta V(kL)^{-1/4}$
appropriate for fast modes.

For solar flare conditions $\beta_p< 0.1$, 
and $\beta_A=v_A/c\sim 10^{-2}$. Assuming $M_A\leq 1$ the above cascade times are about few
seconds at the injection scale but much shorter at shorter scales.

Alf\'ven and slow 
modes are inefficient in scattering energetic particles 
(Chandran 2000, YL02). YL02
identified isotropic fast modes as the dominant scattering agent. 
It is also possible to show (see YL03) that fast modes
are the dominant mechanism for acceleration of particles via {\it
resonant} interaction. Fast modes also dominate the acceleration of particles
through non-resonant interaction (Cho \& Lazarian 2005b). Consequently in 
what follows we will
concentrate on the damping of the fast modes.

\section{DAMPING RATE OF TURBULENCE}

The second important process determining the spectrum of turbulence is its 
damping rate.
Damping becomes important whenever the damping time $\Gamma^{-1}(k)$
becomes comparable to or shorter than the  cascading time $\tau_{cas}(k)$.
As we shall show below, for solar flare conditions, 
the damping time is longer than the cascade time at 
large scales but  decreases faster with decreasing scale and becomes dominant
above the critical wave vector where  $\Gamma(k_c)\tau_{cas}(k_c)=1$.
In this section we derive the damping rate and the critical wavenumber $k_c$.
We first describe the damping by the background thermal plasma.

\subsection{Thermal Damping}

In fully ionized plasma, the damping can be divided into two parts:
collisional and collisionless with their regimes of relevance
determined by the ratio of the turbulence scale  and the 
Coulomb collision mean free path of the background plasma
(Braginskii 1965),
\be
\lambda_{\rm Coul}\sim 9\times 10^{7}{\rm cm}\left(\frac{T}{10^7{\rm K}}\right)^{2}\left(\frac{10^{10}{\rm cm}^{-3}}{n}\right).
\label{meanfp}
\ee
Viscous damping is important for scales $l >\lambda_{\rm Coul}$, so that it 
can play a role between the injection
scale $<L\sim 10^9$ cm and $\lambda_{\rm Coul}\sim 10^8$ cm. For smaller scales, 
$k\lambda_{\rm Coul}>1$, the damping rate is determined by less efficient  collisionless 
processes.

\subsubsection{Viscous Damping}

The viscous damping rate is derived in Appendix A where we show that for low beta
plasma of interest here we have
\be\label{visc}
\Gamma_{vis}(k, \theta)=0.13\tau_0^{-1}\beta_p^{1/2}(kL)(k\lambda_{\rm Coul})\sin^2\theta, 
\,\,\,\, {\rm for} \,\,\,\, k\lambda_{\rm Coul}<1.
\ee
By equating $\Gamma^{-1}(k)$ from above equation to
$\tau_{cas}$ in equation (\ref{tcasfast}) we obtain the critical scale or wavevector
\be
\label{kcritvisc}
k_c\lambda_{\rm Coul}=3.9(M_A/\sin \theta)^{4/3}(\lambda_{\rm Coul}/L\beta_p)^{1/3}.
\ee
For $M_A\sim 1$ and $\beta_p<0.1$ the last two terms are greater than one indicating that
the critical scale is less than the Coulomb mean free path where this damping rate is 
not valid; $k_c\lambda_{\rm Coul} < 1$ if $\sin \theta > 2.8 M_A$, so that  viscous damping 
could be marginally important for $M_A<0.3$.

\subsubsection{Collisionless Thermal Damping}

The nature of collisionless damping is closely related to the radiation
of charged particles in magnetic field. Charged particles
can emit plasma waves through acceleration (cyclotron radiation) and
Cerenkov effect, and can also absorb the radiation under the same condition and cause damping
of the waves
(Ginsburg 1961). For example the gyroresonance with thermal ions causes the damping of the
modes with frequencies close to the ion-cyclotron frequency (Leamon, Mathaeus \& Smith 1998). 
The particles can 
also be accelerated either by
the parallel electric field (Landau damping)
or the magnetic mirror associated with the comoving compressible modes 
under the Cerenkov condition $k_{\parallel}v_{\parallel}\simeq \omega$, known as transit time 
damping, or TTD for short. Because head-on collisions are more frequent than trailing 
ones,  the energy is transfered from waves to particles. 

For small amplitude waves, particles should have parallel speed comparable to wave phase velocity
to be trapped in the moving mirrors. This gives rise to the above Chernenko condition. 
For a thermal plasma, this requires thermal speed $v_{th}\sim v_{ph}$ or a plasma beta
of oder unity. At lower values of  $\beta_p$ the fraction of particles satisfying this condition 
is smaller which means that the damping rate decreases with decreasing $\beta_p$. 
The damping rate 
of fast mode with frequency $\omega=kv_A$ and  $\beta_p< 1$ can be written as  

\be
\Gamma_{th}(k, \theta)  =  \Gamma_0\times \left[\exp\left(-\frac{\delta}{\beta_p\cos^2\theta}\right) +  \frac{5}{\sqrt{\delta}}\exp\left(-\frac{1}{\beta_p\cos^{2}\theta}\right)\right]g(\theta),
\label{thermdamp}
\ee
where, as stated above $\delta=m_e/m_p$ is the  electron to proton ratio and we have
defined a characteristic damping rate
\be\label{gamma0}
\Gamma_0 \equiv \frac{\sqrt{\pi\beta_p\delta}}{2\tau_0}(kL)\frac{\sin^{2}\theta}{\cos\theta}.
\ee
This damping rate, without the last term $g(\theta)$ and valid for 
$\theta\gtrsim \sqrt{\omega/\Omega_p}$, coincides with the one in
Ginsburg (1961).%
\footnote{Note that this expression differs from the one in Ginsburg by a factor of 2. This is 
because we are concerned with the energy instead of wave amplitude damping.}
In the square bracket the first term represents the contribution 
from electrons, and the second term is due to protons. 
The function $g(\theta)$ for $\theta \ll 1$ is 
\be\label{gtheta}
g(\theta)=\frac{1}{2}\left(1+\frac{\theta^{2}}{\sqrt{\theta^{4}+4\omega^{2}/\Omega_{i}^{2}}}\right)\rightarrow
\cases{1, &if $1\gg \theta\gg \sqrt{\omega/\Omega_p}$;\cr
0.5, &if $\theta\ll \sqrt{\omega/\Omega_p}\ll 1$.\cr}
\ee
derived by Stepanov (1958) extends the relation to small angles where the damping rate
decreases by a factor of two.
(In the limit of $\theta=0^o$, there are no compressions, fast modes are degenerate with Alfv\'en
modes.) For $\beta_p\lesssim 0.1$ and sufficiently large $\theta$, the damping due 
to electrons dominates and the damping rate can be written in a simple form
\be
\Gamma_{th}(k, \theta)=\Gamma_0\exp[-\delta/(\beta_p\cos^2\theta)], \,\,\,\,{\rm for} \,\,\,\, k\lambda_{\rm Coul}>1,
\label{thelecdamp}
\ee
where we have ignored the correction $g(\theta)$ at small angles. 
Note the similarity of this relation with that for 
viscous damping; the main difference is the absence of the extra term $(k\lambda_{\rm Coul})$ 
in equation (\ref{visc}). One can then combine the two expressions to obtain an approximate damping
rate valid at all scales
\be\label{totdamp}
\Gamma_{tot}(k, \theta) \simeq \Gamma_{th}(k, \theta)/(1 + \zeta^{-1}),\,\,\,\, {\rm for}\,\,\,\,\zeta=6.3\cos\theta (k \lambda_{\rm Coul}),
\ee 
where we 
have deleted  the exponential part in equation (\ref{thelecdamp}) which is 
equal to one except at small range of angles near $\pi/2$ 
(i.e. $\cos \theta < .023/\sqrt \beta_p$). A more 
accurate expression is obtained if one divides $\zeta$ by the  square bracketed term and 
$g(\theta)$ in
equation (\ref{thermdamp}).

In Figure (1) we compare the cascading time  with the damping time
($\tau_d=1/\Gamma_{th}$) at different scales for
$\beta_p=0.01$ and different angles (right panel), and for $\theta=45^o$ at three values
of $\beta_p=0.001, 0.01$ and 0.1 (left panel), corresponding to magnetic fields of 
$B\sim 600, 180$ and 60 G, respectively (see eq[\ref{beta}]). We use 
typical solar flare values; temperature $T=10^7$ K, density $n=10^{10}\rm cm^{-3}$, 
and we set $M_A=(\delta V/v_A) =0.3$. 
The angular dependence enters this damping by two competing 
factors. In general the damping increase with $\theta$ because magnetic compression increases
so that more particles can be trapped and interact with the waves. However, when $\theta$ approaches 
to $90^o$, {\it i.e. for quasi-perpendicular propagation}, 
most thermal particles will not be in resonance with the fast mode 
waves in a low $\beta_p$ medium, which explains the decrease of damping in this regime. 

\begin{figure}\label{times}
\plottwo{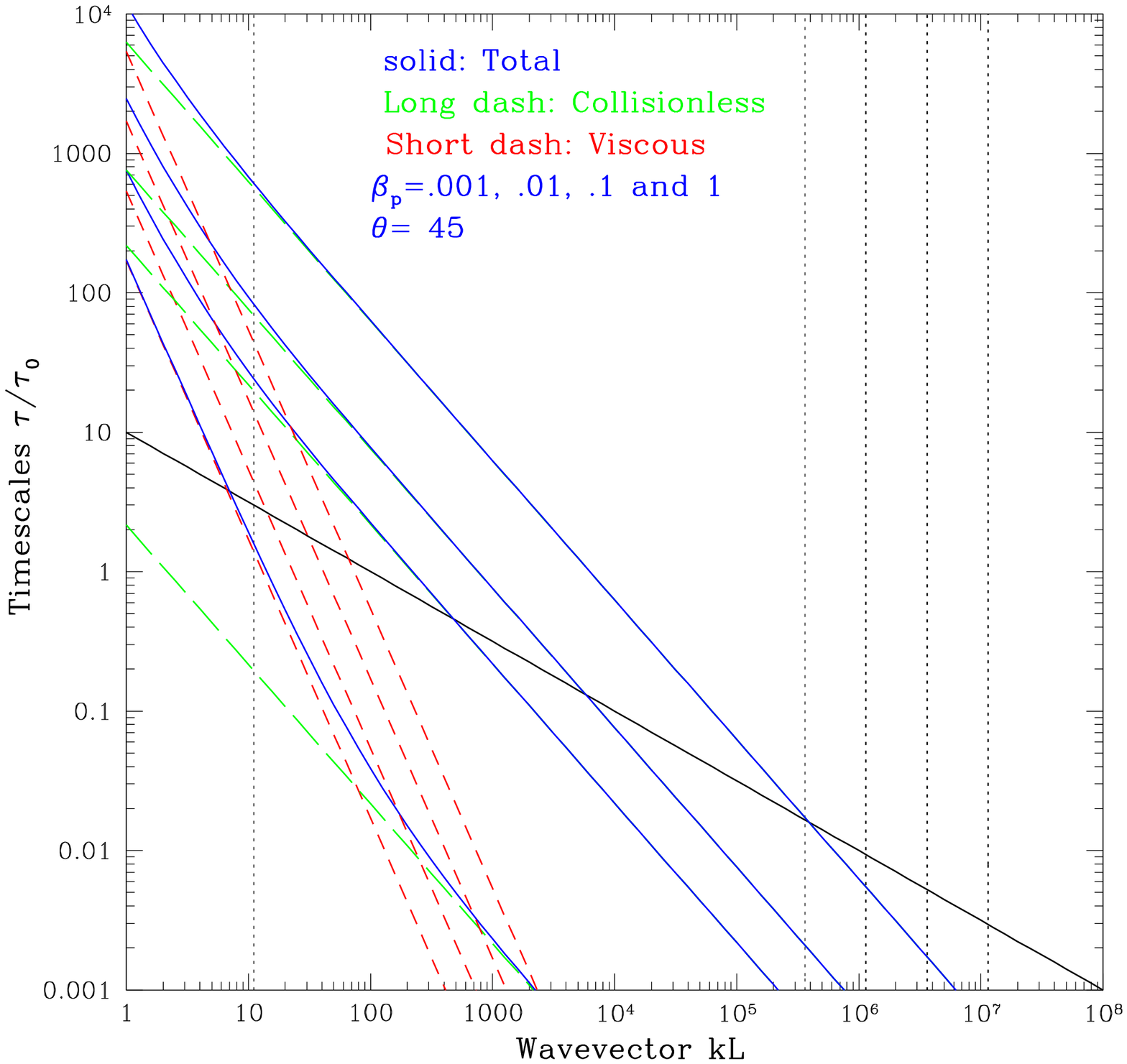}{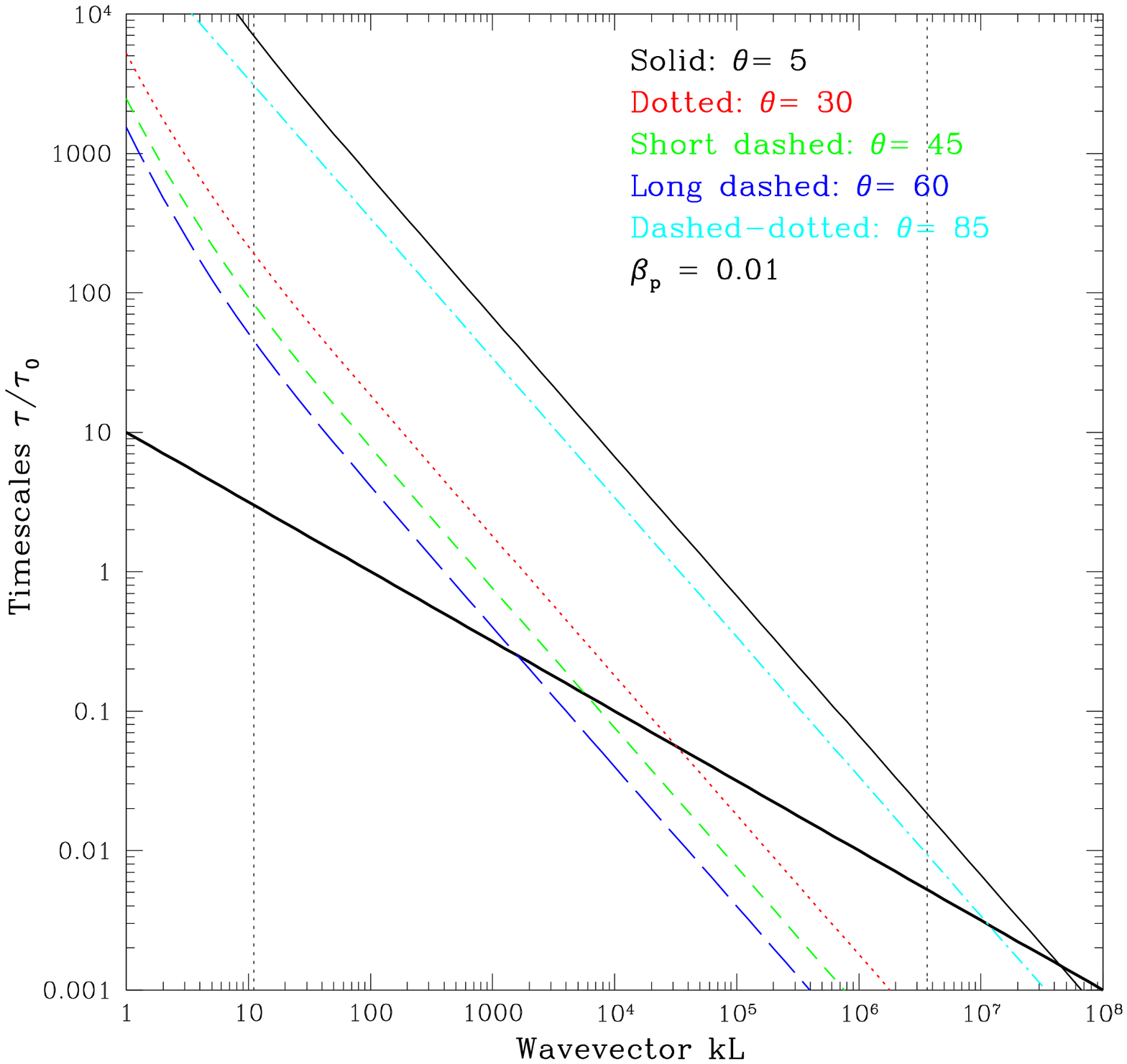}
\caption{Variation with wavevector $k$ (in units of injection scale $L^{-1}$) 
of the cascade and damping timescales (in units of $\tau_0=L/v_A$)
of fast modes at $\theta=45^o$ for different values of $\beta_p$ ({\it left panel}) and 
for various $\theta$ for $\beta_p=0.01$ ({\it right panel}).  On the left panel the short dashed, 
long dashed and solid lines 
are for viscous damping, collisionless damping and the total damping, respectively.
Note the steepening of the total damping at 
low values  $kL<L/\lambda_{\rm Coul}$ (represented by the left vertical dotted line) 
due to emergence of viscous damping.
The critical damping wavevector $k_c$ is given by the intersection of 
the two time scales, which as shown in Figure (2) depends on the angle $\theta$. 
The vertical dotted lines at the right side show the scale of the proton thermal 
gyro radius $L/r_{g,p}$}
\end{figure}

By equating the collisionless damping time from equation (\ref{thelecdamp}) 
with the cascade time in  equation (\ref{tcasfast}), we attain the critical wave vector
\be
k_cL = \frac{4M_A^4\cos^2\theta}{\pi\delta\beta_p\sin^4\theta} \exp\left(\frac{2\delta}{\beta_p\cos^2\theta}\right).
\label{kcritel}
\ee
The variation of $k_c$ with angle  for the thermal collisionless damping using the 
exact expressions is shown in 
Figure (2).
As evident the damping scale given by equation (\ref{kcritel}) varies considerably especially when 
$\theta\rightarrow 0$ and $\theta\rightarrow 90^o$, where it becomes smaller than the 
proton gyro radius (shown by the dashed line). Note also that for  $\beta_p<0.1$ the damping 
scale is larger than the collision mean free path (or $k_c\lambda_{\rm Coul}>1$, shown
by the dashed-dot horizontal line), except
for few degree around $\theta=85^o$, which is within the range of its validity. 
The specific range of the $\theta$ where the relations breakdown depends on the plasma $\beta_p$.

This describes the well known fact  that at large $k$ vectors ({\it i.e.} small scales) 
the turbulence will be
very anisotropic. Only 
quasi-parallel and quasi-perpendicular modes survive at such large wavevectors 
(and corresponding frequencies) which are needed in the 
acceleration of low energy particles. However, as turbulence undergoes  cascade and/or 
waves propagate in a turbulent medium the character of this anisotropy changes because the
angle $\theta$ is changing due
to the randomization of wave vector $\bf k$ 
and the wandering of the magnetic field lines discussed next.

\begin{figure}
\plotone{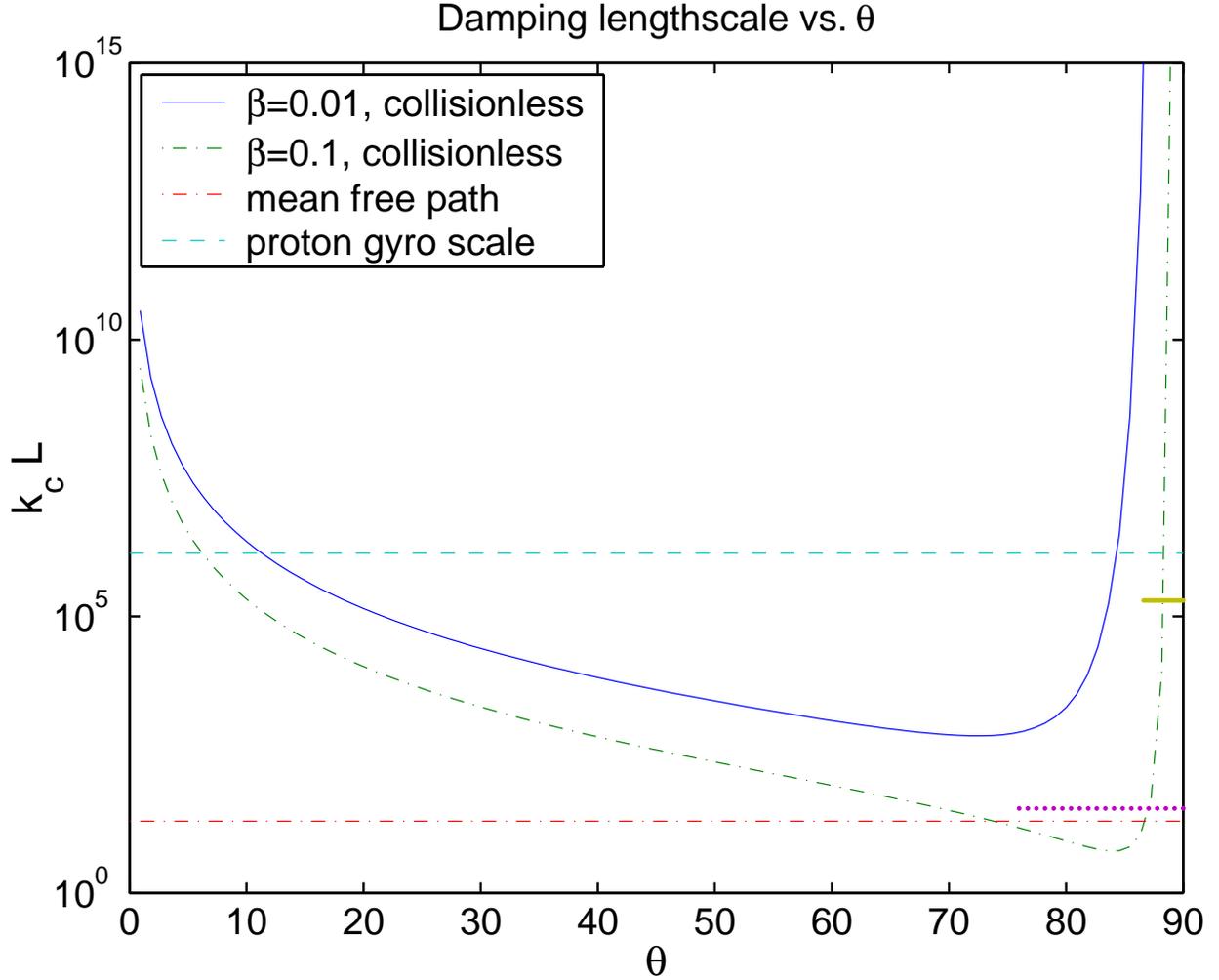}
\caption{Normalized cutoff scales of turbulence due to thermal damping vs. 
the angle $\theta$ between {\bf k} and {\bf B}  obtained by equating turbulence cascading  
and damping rates from 
equations (\ref{tcasfast}) and (\ref{thermdamp}). The horizontal dash-dotted and dashed 
lines represents the 
scales of the mean free path ($L/\lambda_{\rm Coul}$), the dividing line between 
collisional and collision less damping, and the thermal proton gyro radius 
($L/r_{g,p}$), the limit of applicability of MHD, respectively. 
The dotted lines show the effect of field wandering and ${\bf k}$
vector randomization at angles near $90^o$ for $\beta_p=0.1$. The wanderings are ineffective 
at $\beta_p<0.05$. Note that the critical scale is almost always smaller
than the mean free path so that the viscous damping  can play only a marginal role (see text).}
\label{kcrit}
\end{figure}

\subsubsection{Damping Anisotropy}

The damping of fast modes described above is valid for small perturbations 
and for 
a uniform background magnetic field.
A more realistic setting for damping in turbulent media, which is
based on better understanding of turbulent cascade (Cho \& Lazarian 2002, 2003)
and magnetic field wandering (Lazarian \& Vishniac 1999, 
Lazarian, Vishniac \& Cho 2004),  
is discussed in YL04 for the case of interstellar medium. Here we apply this approach for turbulence in
solar  flares.

For fast mode cascade, the non-linear 
cascading occurs by interaction of wave packets  that are 
collinear (see review by Cho \& Lazarian 2004), 
$\omega = \omega_1 + \omega_2$ 
and ${\bf k}={\bf k}_1 + {\bf k}_2$, and the final wave is parallel to the initial ones. 
However, the energy conservation is true within an  uncertainty condition 
$\delta \omega\sim \tau_{cas}^{-1}\ll \omega$, which will give rise to an orthogonal 
wavevector component.
It is easy to show then that this uncertainty in frequency will yield a small transverse component 
$\delta k$ related to $\delta \omega$ as $\delta \omega \sim V_{ph} \delta k (\delta k/k)$. 
For $\delta\omega/\omega\sim 1/(\omega\tau_{cas})\ll 1$ the angle $\delta \theta$ between the final 
and 
initial wavevectors will be small. Combining this with equation (3) we  get for a given Alfvenic Mach number
$M_A$ the following expression (YL04)
\be
\tan \delta \theta\simeq  \delta k/k\simeq (\delta \omega/\omega)^{1/2} \simeq M_A (kL)^{-1/4} ,
\label{dthetak}
\ee 

It is easy to see that, as in interestellar medium case, the field line 
wandering in the case of solar flares is mainly caused by the shearing due to 
Alfv\'en modes.  According to Lazarian \& Vishniac (1999)\footnote{A typo of numerical coefficient has been corrected.} the field line diffusion along 
and perpendicular the mean field provides
\be
\tan \delta\theta_\parallel \simeq M_A^2\left(\frac{z}{27L}\right)^{1/2},\,\,\,\,\,
\tan \delta\theta_\perp \simeq M_A\left(\frac{r}{L}\right)^{1/3} 
\ee
where $z$ and $r$ are the distances along and perpendicular to the mean field direction. 
During one cascading time (equation \ref{tcasfast}), the fast modes propagate a distance 
$\tau_{cas} = \tau_0/(M_A^2 \sqrt {kL})$ and see an angular deviation 
\be
\tan \delta \theta \simeq \sqrt{\tan^2\delta \theta_\parallel+\tan^2 \delta\theta_\perp} \simeq \sqrt{\frac{M_A^2\cos\theta}{27(kL)^{1/2}}+\left(\frac{M_A^2\sin^2\theta}{kL}\right)^{1/3}}
\label{dthetaB}
\ee

Note that at the largest scale $kL=1$ the randomization is of the order of $M_A < 1$
\footnote{It is useful, in general,
to define the domain of MHD turbulence as starting from Alfven Mach number of unity (see discussion in 
Cho \& Lazarian 2005a). This condition is automatically satisfied for strongly magnetized 
turbulence in solar flares.}
and decreases slowly with $k$; $\delta \theta$ scales as $(kL)^{-1/4}$ at 
small angles or quasi-parallel modes and  as $\sim (kL)^{-1/6}$ at other 
angles. This 
shows that randomization decreases with the decrease of the scale.

While for the processes of scattering and acceleration of particles by 
fast modes randomization is important (Yan \& Lazarian 2004), it does not 
always
result in tangible changes of the overall picture of damping. For instance,  
combing equations (\ref{kcritel}) and (\ref{dthetaB}) 
we can show that for quasi-parallel modes (i.e. $\theta \sim 0$) 
at the critical 
wavevectors $k_c$ the randomization angle is rather small, i.e. 
$\delta \theta \sim 10^{-3}M_A\beta_p^{1/2}$.  
For other angles $\delta \theta <(\beta_p/30)^{2/3}$ and is still small but may not be negligible.
In particular,
for the quasi-perpendicular modes, we can make an estimate of the presence of field line
wandering on the damping truncation scale by evaluating the average of the damping rate in 
equation (\ref{thermdamp})  over the small range $\pi/2-\delta\theta$ to $\pi/2$,  where  
from equation (\ref{dthetaB}), $\delta\theta\sim M_A^{2/3}(kL)^{-1/6}$. For $\delta \theta \ll 1$
we can define $\alpha=\pi /2 - \theta$ and use the approximations  
$\sin\theta=1$ and $\cos\theta=\alpha$ 
 so that the average value of the damping rate given in equation 
(\ref{thermdamp}) is roughly given by 
\be\label{averate}
\langle\Gamma_{th}\rangle=\frac{\sqrt{\pi\beta_p\delta}(kL)}{2\tau_0\delta\theta}\int_0^{\delta\theta} d\alpha\exp{-\delta/(\beta_p\alpha^2)}/\alpha=\frac{\sqrt{\pi\beta_p\delta}(kL)}{4\tau_0\delta\theta}{\rm E}_1\left(\frac{\delta}{\beta_p\delta\theta^2}\right).
\ee
Equating this with $\tau_{cas}$ from equation (\ref{tcasfast}) we get $x^2{\rm E}_1(x)=A$ with 
$x=(\delta/\beta_p)(kL/M_A^2)^{1/3}$ and 
$A=(4/\sqrt\pi)M_A^{-1}\delta^{3/2}\beta_p^{-5/2} = (.025/\beta_p)^{-5/2}$,
where we set $M_A=0.3$.
For $\beta_p=0.1$ this relation is satisfied for $x\sim 0.15$ or 
$k_cL=M_A^6(x\beta_p/\delta)^3\sim 20$. This scale is shown by the horizontal dotted line in Figure (2).
This means that modes in the cone near $90^o$  get damped above this scale 
due to randomization of $\delta\theta$.
However, 
because $x^2{\rm E}_1(x)\lesssim 0.22$ there is no solution for $\beta_p < 0.05$ indicating 
that the field wanderings do not reduce the damping scale at $90^o$ 
for highly magnetized plasmas. 

These results are only rough estimates. A more detailed study of the 
issue is more appropriate in the context of particle acceleration in solar
flares and will be done elsewhere.

\subsection{Nonthermal damping}

The wave damping rate calculated above assumes that the energy lost by waves with spectrum 
$W({\bf k})$ goes into heating the plasma and that the plasma maintains its 
Maxwellian distribution, presumably via Coulomb collisions in a timescale of 
$\tau_{\rm Coul}\sim \lambda_{\rm Coul}/c_S$, where $c_S$ is the sound speed.
This requires a longer damping time;
$\Gamma^{-1}_{th}>\tau_{\rm Coul}$. As stated above we are interested 
in a collisionless plasma with $\tau_{\rm cas}\ll\tau_{\rm Coul}$. This combined with the fact that
damping is important when $\Gamma^{-1}_{th}<\tau_{\rm cas}$ implies that the above condition
is not satisfied and some particles get accelerated to energies much higher than $k_BT$.
Because $\tau_{\rm Coul}$ decreases with energy fairly rapidly it can be 
shorter that the other times  at low energies, the particle spectrum there will be 
approximately Maxwellian. 
As shown in PL04 solution of the particle kinetic equation (1) with a given background 
thermal plasma and an assumed spectrum of turbulence does lead to a particle
spectrum consisting of a quasi-thermal part and 
a {\it nonthermal tail} with dividing energy roughly where $\tau_{Coul}$ is equal to the 
acceleration time $\tau_{ac}\simeq E/A(E)$. 
Thus, we need to also consider damping of the 
waves by the nonthermal tail. As mentioned at the 
outset one must carry out this 
self-consistently by solving the coupled wave-particle kinetic equations which is beyond
the scope of this paper. Here we derive the damping rate due to electrons and protons  
with a total density of $N_0$, a power law
energy spectrum, $N(E)=N_0(a-1)(E/E_0)^{-a}/E_0$ for $E>E_0$,   
and isotropic pitch angle distribution.  

As can be
surmised the calculations of the particle diffusion coefficient 
$D_{pp}\equiv\langle \frac{\Delta p \Delta p}{\Delta t}\rangle$, or
the acceleration rate $A(E)/E$ and
the damping rate $\Gamma_{nonth}({\bf k})$ are intimately connected. Let us represent 
the transition rate (integrated over the particle pitch 
angle cosine $\mu$)  of the interaction between a wave ${\bf k}$ and a particle with 
energy $E$ by $\sigma ({\bf k}, E)$. The Fokker-Planck diffusion coefficients $D(p)$, which is $D_{pp}$
integrated over $\mu$, is

\be\label{diffusion}
D(p)=
\int_{-1}^1 d\mu D_{pp}= \int_{k_{min}}^\infty d^3kW({\bf k})\sigma({\bf k}, E).
\ee
From this we can get the rate of systematic energy gain by particles (PL04):
\be\label{accrate}
A(E)=\frac{d [v p^2D(p)]}{4p^2d p}=\int_{k_{min}}^\infty d^3kW({\bf k}) \Sigma({\bf k}, E),
\ee
which also means that 
\be\label{sigma}
\Sigma({\bf k}, E)=\frac{d [v p^2\sigma({\bf k}, E)]}{4p^2d p}. 
\ee
Because the energy lost by turbulence 
${\dot {\cal W}_{nonth}}\equiv \int \gamma_{nonth}({\bf k})W({\bf k})d^3k$ 
to nonthermal particles must be 
equal to the energy gain by these particles, ${\dot {\cal E}}=\int A(E)N(E)dE$, the damping rate is 
given by 
\be\label{damprate}
\Gamma_{nonth}({\bf k})= \int_{E_0}^\infty dE N(E)\Sigma({\bf k}, E).
\ee

The transition rate $\sigma$ for  a particle with gyrofrquency $\Omega$ and pitch 
angle cosine $\mu$ interacting with
a wave of frequency $\omega$ and wavevector $k$
is determined by the resonant condition
\be\label{resonance}
\omega - k\cos\theta v\mu =n\Omega/\gamma, \,\,\,\,n=(0, \pm 1, \pm 2, . . .).
\ee
This rate can be expressed as 
sum  over $n$ of squares of Bessel functions $J_i^2(k_\perp v_\perp\gamma/\Omega)$ 
with $i=n, n-1 {\rm\, or\,} n+1$ and $v_\perp=v(1-\mu^2)^{1/2}$. 
(For details see Pryadko \& Petrosian 1999.) For waves propagating parallel to the $B$ field
$k_\perp=0$ and only $J_0(0)\neq 0$ and $n=\pm 1$ ($n=0$ term also vanishes). 
The resonance condition then requires $k_{res}\sim \Omega/v_\parallel \sim r^{-1}_{g}$. 
Given the parameters we adopt here $r_g\sim 1$ and 50 cm for thermal electrons and protons,
respectively, which is certainly beyond the MHD regime. Only quasi-parallel propagating waves 
cascade to such small scales without undergoing thermal damping and can contribute to acceleration 
of low energy particles (see Fig. 2)

\begin{figure}
\includegraphics[ width=0.45\textwidth]{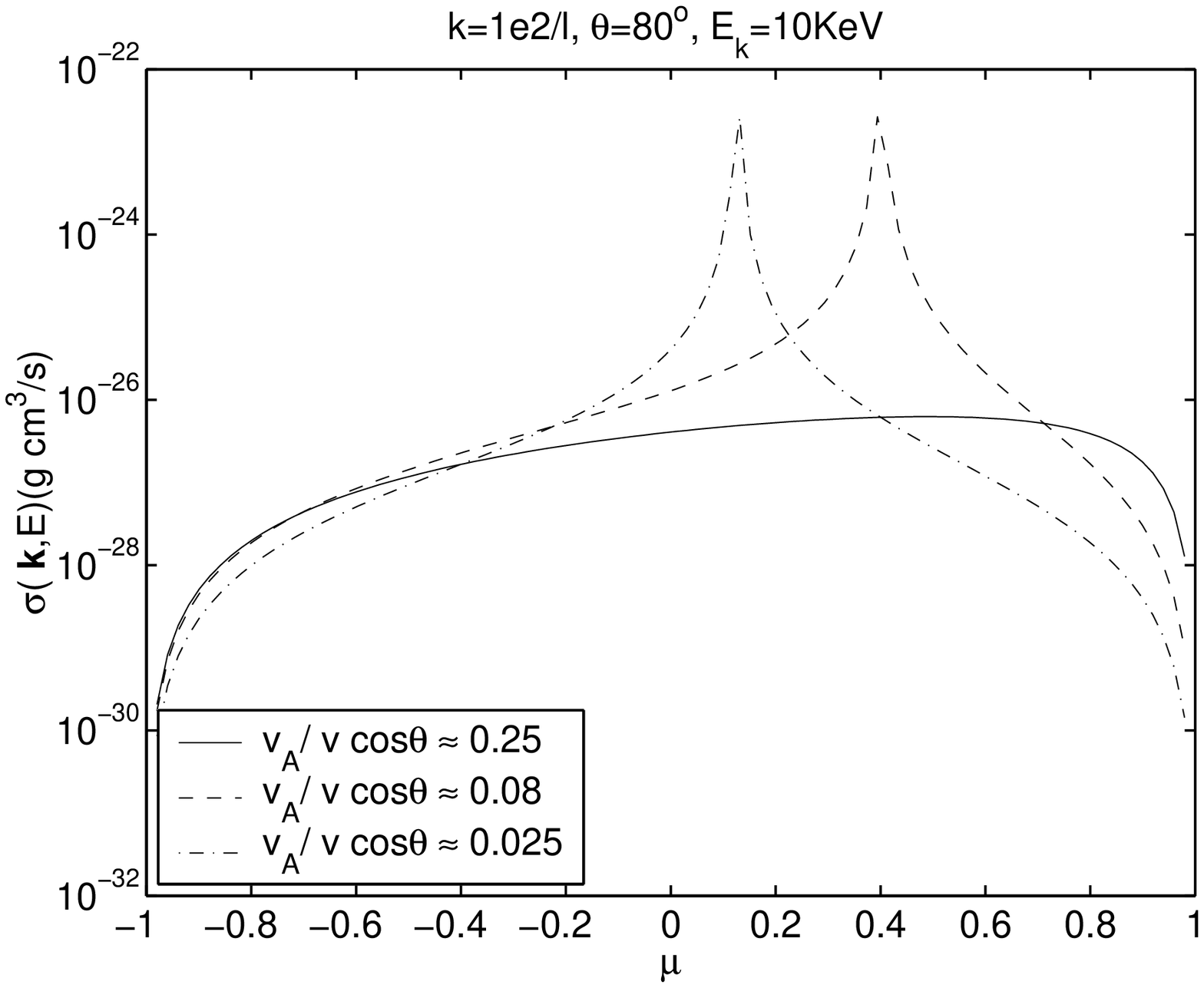}
\includegraphics[ width=0.45\textwidth]{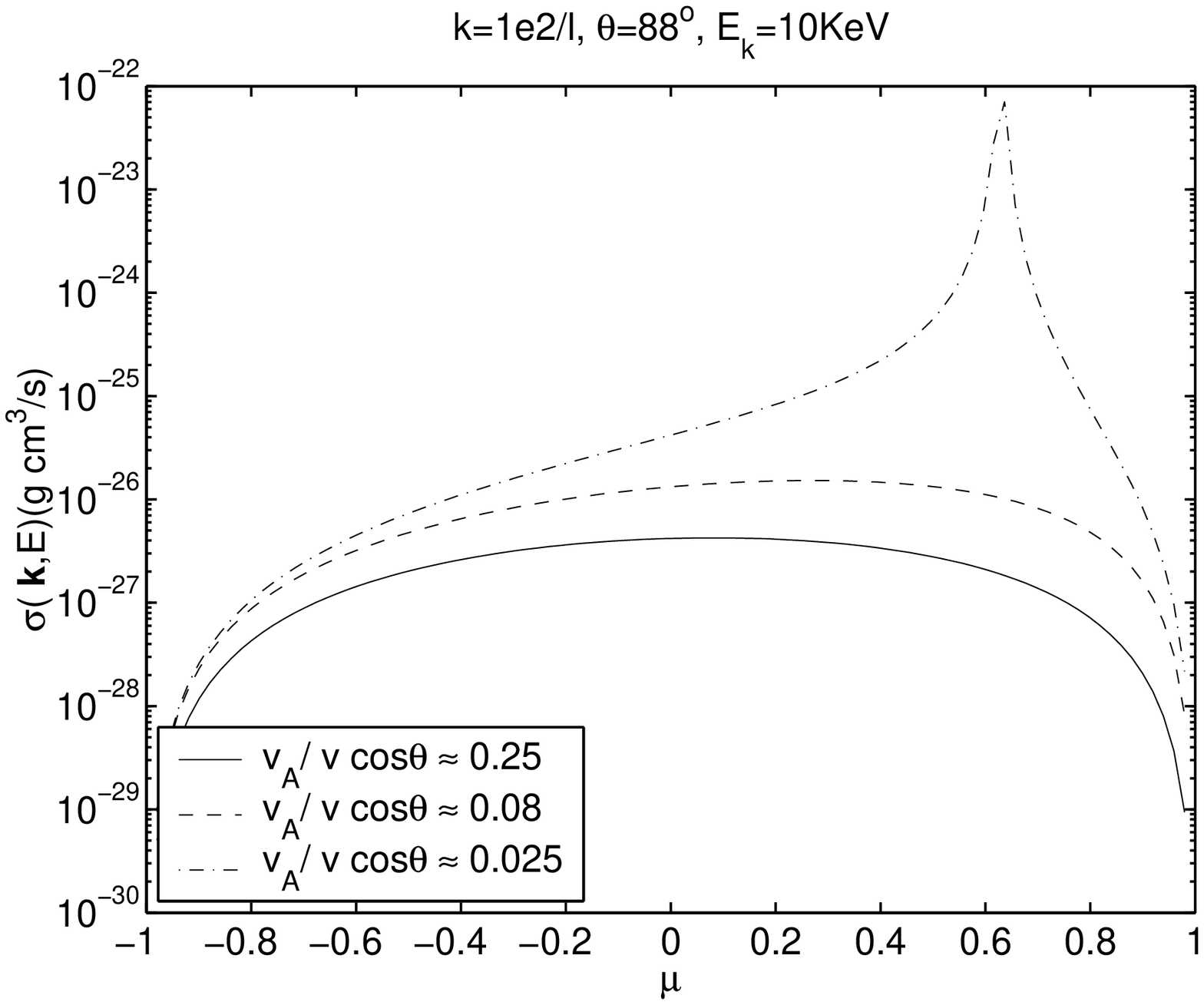}
\includegraphics[ width=0.45\textwidth]{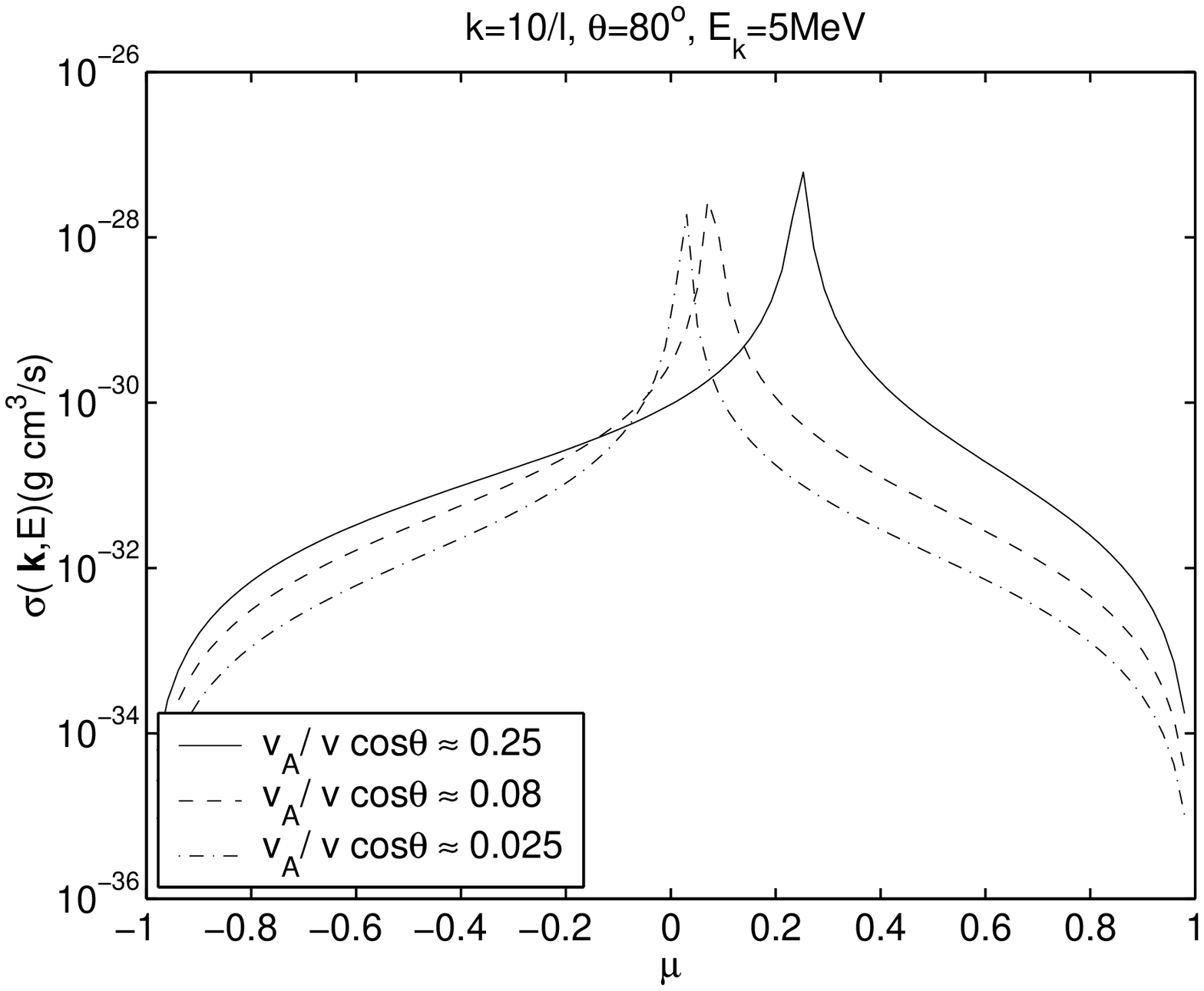}
\hspace{1.4cm}
\includegraphics[ width=0.45\textwidth]{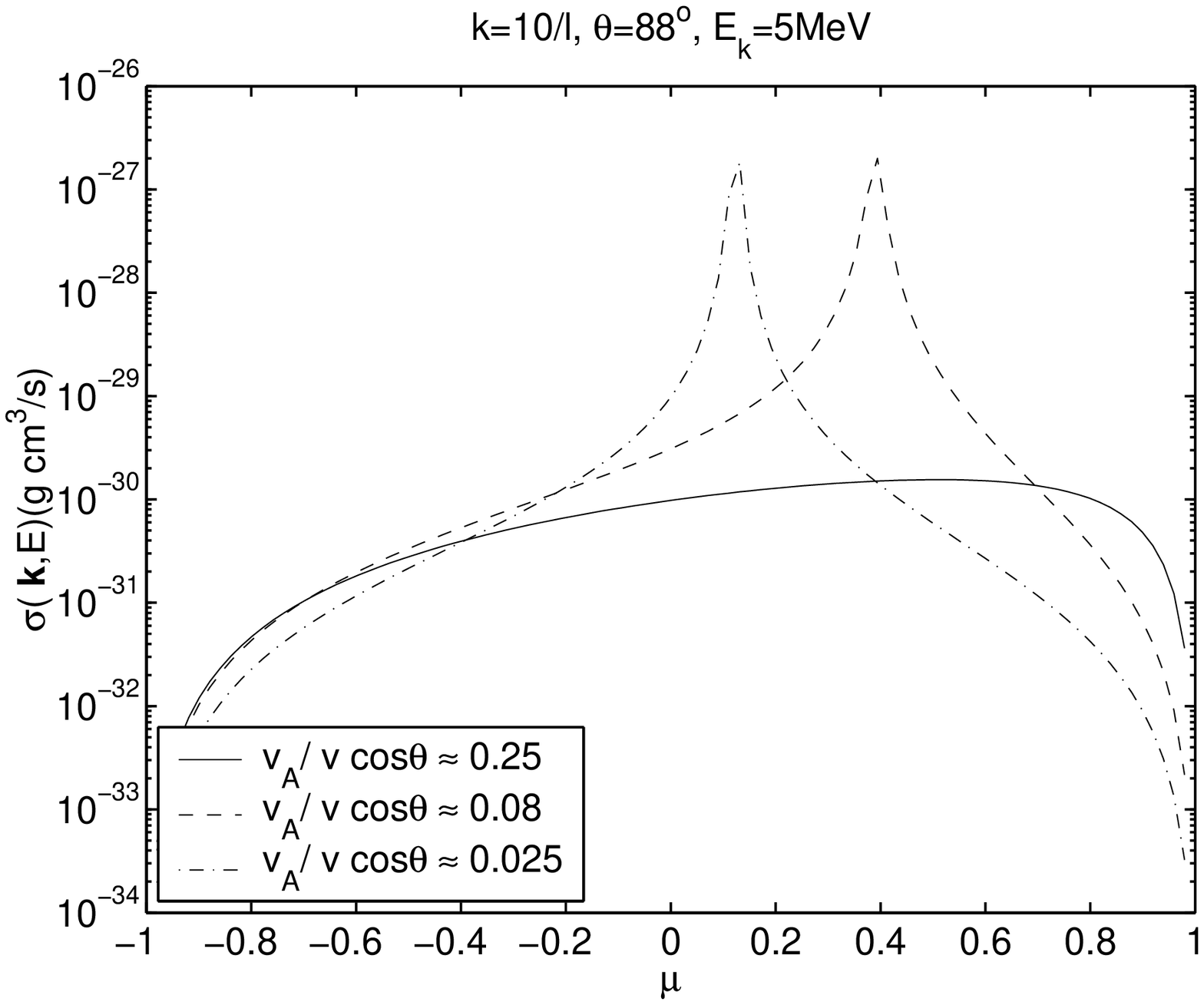}
\caption{The dependence of the diffusion coefficient on 
$\mu$ from equation \ref{ttdrate} for $\theta=80^o$ (left panels) and $\theta=88^o$ (right panels),
and for particle kinetic energies of 10keV (top panels) and 5MeV (bottom panels). 
Most resonance can be approximated by $\delta$ function 
(see text). In 
those cases $v\cos\theta < v_A$, the resonance condition is not  satisfied and therefore there 
is relatively very weak interaction rate.}
\label{integrand}  
\end{figure}

For obliquely propagating waves (fast or Alfv\'{e}n) things are more complicated 
and all $n$'s could contribute. However, these waves are damped at scales  much larger than that 
required for the above resonant condition. Except for  $n=0$ or the {\it transit time damping} 
(or {\bf TTD} for short) mode 
which happens at all scales with the resonant condition $v\mu = v_A/\cos\theta$. 
TTD is resonant interaction with parallel magnetic mirror force.  
Thus, in what follows we consider this process 
with the transition rate

\be\label{ttdrate}
\sigma({\bf k}, E)=
\frac{\Omega m^2}{2nm_p}\int_{-1}^1 d\mu (1-\mu^{2})\left( \frac{\tau_{cas}^{-1}}{\tau_{cas}^{-2}+(k_{\parallel}v\mu -
\omega)^2} \right)J_1^2(x), \,\,\,\,\,x\equiv\frac{\gamma k_\perp v_\perp}{\Omega},
\ee
where $n$ is the density of protons (which we take to be equal to that of electrons), and
The resonance
function in the large parenthesis is produced by 
integration over time. In general, the width of this function 
$\Delta\mu=\tau_{cas}^{-1}/k_\parallel v=M_A^2(kL)^{-1/2}(v_A/v\cos\theta)\ll 1$  because 
usually  Alfv\'en Mach number $M_A< 1$ 
and  because damping normally becomes important at scale $k^{-1}\ll L$.
(This is not true only for nearly 
perpendicular propagation.) Then the resonance function can be 
approximated by a delta function, i.e., 
$\tau_{cas}^{-1}/(\tau_{cas}^{-2}+(k_{\parallel}v\mu -\omega)^2)\rightarrow \pi 
\delta(k_{\parallel}v\mu -\omega)=\pi\delta(\mu-\mu_{res})/k_\parallel v$, where 
$\mu_{res}=v_A/v\cos\theta$. In 
Figure (\ref{integrand}), the integrand of $\sigma({\bf k},E)$ is plotted versus $\mu$ 
for some interesting cases. As evident, when the  resonance condition is satisfied, {\it i.e.}
$v_\parallel< v_A\cos\theta$, the transition rate $\sigma$ peaks sharply by many orders of magnitude
so that it can be well represented by a delta function.

With this simplification the integration over $\mu$ can then be 
carried out easily. Then from the relation $J_1'=(J_0-J_2)/2$ and equations 
(\ref{ttdrate}), (\ref{damprate})
and (\ref{sigma}) we can obtain  the damping rate 
\begin{eqnarray}\label{drate}
&&\Gamma_{nonth}({\bf k})=\frac{\pi}{8} \frac{\Omega^2
 m}{nm_p ck\eta}\left(1-\frac{\beta_A^2}
{\eta^2}\right)\int_{E_0}^{\infty} 
dE N(E) \Theta (E - E_c)/(\beta\gamma) \nonumber\\
&&\left[2J_1^2(x)+xJ_1(x)\right(J_0(x)-J_2(x)\left)\right],\,\,\,\,\, 
x\equiv\beta\gamma ck \Omega^{-1}\sqrt{1-\eta^2} \sqrt{1-\beta_A^2/(\beta\eta)^2}.
\end{eqnarray}
Here 
we have defined $\eta=\cos\theta$ and $E_c/mc^2=(1/\sqrt{1-\beta_A^2/\eta^2}) - 1$, 
and $\Theta(x)=1\,\, {\rm for}\,\, x>1$, and zero otherwise,  is the Heaviside step function.

{\bf Damping by Electrons:} For electrons $kc/\Omega_e\sim 10^{-8}(kL)(100 {\rm G}/B)\ll 1$ 
for the relevant scales $k<k_c$. Therefore, except for extreme relativistic
electrons the variable $x\ll 1$ and we can use the first order 
approximation $J_{n}(x)\simeq(x/2)^{n}/n!$ for the Bessel functions. Then the terms 
in the square brackets in the above equation can be approximated as $x^2/2$  to give 

\be\label{erate1}
\Gamma_{nonth}({\bf k})=\frac{\pi}{8}\frac{\delta}{n\beta_A}\frac{kL}{\tau_0}\left(1-\frac{\beta_A^2}{\eta^2}\right)
\left(\frac{1-\eta^2}{\eta}\right)\int_{E_m}^{\infty} dE N(E)\beta\gamma\left(1-\frac{\beta_A^2}{\beta^2\eta^2}\right),
\ee
where $E_m={\rm max}(E_0, E_c)$.

We first note that if we use a {\it nonrelativisitic Maxwellian}  
distribution, $N(E)=n(2/\sqrt{\pi})(k_BT)^{-3/2}E^{1/2}\exp{(-E/k_BT})$,
this gives the damping rate of
\be\label{eratetherm}
\Gamma_{th}({\bf k})=\Gamma_0\exp\left(-\frac{\delta}{\beta_p\cos^2\theta}\right)
\ee
which is identical to the electron part of the collisionless thermal damping given in equation
(\ref{thermdamp}).

For a {\it nonthermal} distribution we can carryout the integration which leads to a complicated 
expression shown in Appendix B. In the nonrelativistic and extreme relativistic cases the 
result simplifies considerably. For the spectral index $a>2$ most of the contribution 
to the integral comes from the low energies so that when $E_m\ll m_ec^2$ ({\it i.e.} $E_0\ll m_ec^2$ and
$E_c/m_ec^2\simeq\beta_A^2/(2\eta^2)\ll 1$) we can use the {\it nonrelativistic approximation} to obtain

\be\label{eratenonrel}
\Gamma_{nonth}^{nonrel}({\bf k})=\Gamma_0\left(1-\frac{\beta_A^2}{\eta^2}\right)\left[\left(\frac{\sqrt{\pi}(a-1)N_0}{4(a-3/2)n}\right)
\left(\frac{E_0}{k_BT}\right)^{1/2}\right]
\cases{1-\left(\frac{(a-3/2)E_c}{(a-1/2)E_0}\right), &if $E_0\gg E_c$;\cr
\left(\frac{1}{2a-1}\right)\left(\frac{E_c}{E_0}\right)^{3/2-a}, &if $E_0\ll E_c$.\cr}
\ee

The second expression is valid near $\theta=\pi/2$ where $\eta\ll \beta_A/\sqrt{E_0}$. For the solar 
flare conditions we are considering here, $\beta_A\sim 0.007$ and $E_0\sim 0.02m_ec^2\sim 10$ keV, 
this expression is applicable only 
for $\pi/2 - \theta<0.05$ or within $0.3^o$ of perpendicular direction.
Thus, for all angles outside this range we have the first expression which is different from   
the thermal damping rate of equation (\ref{eratetherm})
by the presence of the terms in the square brackets.
The main part here is the ratio
$N_0/n$ of the nonthermal to thermal particle densities and by $(E_0/k_BT)^{1/2}$, {\it i.e.} the ratio
of the mean momenta (or velocities) of nonthermal to thermal electrons. 
In general, this term is less than one and 
damping by thermal electrons is dominant. However, in some large solar flares one requires 
acceleration of a large fraction of the background thermal particles to energies $\gg k_BT$ 
so that the damping 
by nonthermal particles could be significant. 

If $E_0\gg m_ec^2$, we can use the {\it extreme 
relativistic} approximation to obtain

\be\label{erateextrel}
\Gamma_{nonth}^{rel}({\bf k})=\Gamma_0\left(1-\frac{\beta_A^2}{\eta^2}\right)^2\left(\frac{\pi}{32}\right)^{1/2}\left(\frac{N_0}{n}\right)\left(\frac{a-1}{a-2}\right)\left(\frac{E_0}{k_BT}\right)^{1/2}\left(\frac{E_0}{m_ec^2}\right)^{1/2},
\ee
which is valid for all angles, where a $E_c/m_ec^2\sim \beta_A^2/\eta^2\ll E_0/m_ec^2$, 
so that $E_m=E_0\gg m_ec^2$, except for a extremely narrow range of angles give by
$(1-(m_ec^2/E_0)^2\beta_A^2\ll \eta^2 \ll \beta_A^2$.
One can combine the above  two expressions to obtain an approximate relation valid at all energies:
\be\label{allE}
\Gamma_{nonth}({\bf k})=\Gamma_{nonth}^{nonrel}\left(1+\frac{a-3/2}{\sqrt{2}(a-2)}\left(\frac{E_0}{m_ec^2}\right)^{1/2}\right).
\ee
However, it should be noted that the extreme relativistic equation is valid only for scales
$k<\Omega/[c\gamma_0\sin \theta\sqrt{1-\beta_A^2/\eta^2}]$, otherwise the 
approximation used for the Bessel functions breaks down. For most angles, and for 
parameter values adapted here, this means $kl\lesssim 5\times 10^7/\gamma_0$. But for 
$\theta\rightarrow 0$ or $\cos\theta\rightarrow \beta_A$ the above expression would be valid
at much smaller scales or larger values of $kL$. These limitations are also true for the more
general equation (B1) in the appendix.

{\bf Damping by ions:}  For protons (and other heavier ions) the condition $k_\perp 
v_\perp/\Omega_i\ll 1$ is not always satisfied.  Nevertheless, if we  use the small argument 
asymptotic expression $J_{n}(x)\sim(x/2)^{n}/n!$ 
for Bessel functions (as done above for electrons), 
we can get similar estimate for  the damping rate due to interaction with protons. 

For example, for a {\it Maxwellian} proton distribution one can show that the resultant 
damping rate will be same as that for electrons with $m_e\rightarrow m_p$, which means setting 
$\delta=1$ in the equation (\ref{eratetherm}). Aside from the factor of 5 this is identical to 
the contribution by protons to the collisionless thermal damping in equation (\ref{thermdamp}).

Within a similar accuracy we can also estimate the damping rate due to {\it nonthermal protons}.
Ignoring angles near $\pi /2$ for the moment, from equations (\ref{drate}) or (\ref{erate1})
we note that $\Gamma_{nonth} \propto \langle p\rangle N_0$ so the relative importance 
of protons and electrons 
depend on their total number ratios and their mean momenta which will be same as the ratio of the
momenta  
$p_0$  at the low end of the spectrum. In solar flares we deal with nonrelativistic values 
of $E_0=\sqrt{2mp_0}$ 
and much fewer number of accelerated protons compared to electrons. In majority
of flares the ratio of the 
total energies $R\equiv (N_0E_0)_p/(N_0E_0)_e$ (in the observable range 
($E_{0,e}>10$ keV for electrons and $E_{0,p}>10$ MeV for protons)  is much less than one 
and varies from 0.01 to 10 in flares with detectable gamma-ray line emission produced by
the accelerated protons and ions
(Miller et al. 1997, attributed to R. Ramaty \& N. Mandzavidze). 
Thus $\Gamma_{nonth,p}/\Gamma_{nonth,e} = R\sqrt{E_{0,e}/(E_{0,p}\delta)}\sim R$ for the above 
mentioned energies. 
In summary, usually one can ignore the nonthermal damping due to protons relative to electrons 
(as was the case for the collisionless thermal damping)
in most solar flares. But nonthermal damping by protons could 
be more important than that of nonthermal electrons in flares with strong gamma-ray line emission.

 \section{SUMMARY AND DISCUSSION}
 
The study of interactions of plasma waves and turbulence with particles of
magnetized plasma is a complex process  and requires an accurate
formulation of the cascade of the turbulence from a large injection
scale to smaller scales and the damping of the waves by  the
background  thermal particles and those accelerated  into a
super-thermal  power-law  tail arising from these
interactions. Equipped with this knowledge one can then determine the
evolution of the spectrum and angular characteristics of the
turbulence and particles by solving the coupled kinetic equations
(see eq.[\ref{kineq}]). There has been considerable progress in the
understanding of the cascade of turbulence from an injected scale $L$
to lower scales or higher wavevector $k$ and its expected spectral
and angular characteristics. We briefly review these
for magnetically dominated or  low beta plasmas 
($\beta_p\ll 1$), such
as those envisioned for solar flares, and indicate that of the three MHD modes
Alfv\'en, slow and fast, the latter can play a dominant role in
heating and acceleration of plasma particles. The aspect of this
process most relevant to our goals in this paper is the rate of
cascade or cascade time as a function of scale of the turbulence. In
general, the  cascade time $\tau_{cas}$ for all these modes is of order of Alfv\'en
injection scale time $\tau_0\equiv L/v_A$ and decreases as
 $(kL)^{-1/2}$. 

The spectral and angular distribution of the turbulence is further
modified at smaller scales when the damping rate becomes comparable and
larger than the cascade rate. The main goal of this paper is to give a
complete description of the damping process. We 
review the basic processes involved here and present equations
describing the damping rate of the turbulence due to different
mechanisms. We first consider viscous damping  valid on scales larger
than the collision mean free path, or for $k\lambda_{\rm Coul}<1$, with the
damping time scale
$\tau_d \sim \tau_0[\sqrt \beta_p(kL)(k\lambda_{\rm Coul})]^{-1}
\propto k^{-2}$. Because of this rapid decline this damping can become
quickly important and stop the cascade process. This would be the case
for high beta plasmas but for solar flare conditions, this damping
mechanism is applicable in  the narrow range of scales between the
injection scale  $L\sim 10^9$ cm and $\lambda_{\rm Coul} \sim 10^8$
cm, where because of small value of $\beta_p$, $\tau_d>\tau_{cas}$ and can be neglected. 
For smaller scales the damping is produced by
collisionless processes. Here we have described the damping due to 
thermal (Maxwellian) and a nonthermal (power-law)  distributions,
separately.

The thermal damping is dominated by electrons and for most practical
purposes can be approximated as $\tau_d \sim \tau_0[\sqrt
\beta_p(kL)]^{-1} \propto k^{-1}$. Proton contribution to this process
can be important for $\beta_p\cos^2\theta> 0.18$ which
will not be the case for low beta plasmas $\beta_p<0.1$
under consideration here. Here $\theta$ is the angle between the
magnetic field and the propagation direction. The same is true for
other ions. We combine the collisionless damping valid for
$k\lambda_{\rm Coul}>1$ with that for the viscous damping and give a
simple expression valid approximately at all scales. Equating the
damping and cascade times we determine the critical wavevector $k_c$
above which the damping becomes dominant and would cause the spectrum
of the turbulence to steepen. The damping is highly anisotropic and the 
critical wavevector varies considerably
with angle $\theta$, being much larger for quasi-parallel and
quasi-perpendicular propagations. However, we show that this anisotropy
around the perpendicular direction is smoothed out by magnetic field
wanderings caused by the shearing due to Alfv\'en modes for $\beta_p<0.05$. 
The quasi-parallel waves, on the other hand, are not affected by this process
and can survive without damping to scales as small as the particle
gyro radii where the MHD regime breaks down and other plasma and kinetic
effects become important. These parallel propagating waves may be then
the most important agents for acceleration of low energy electrons,
protons and other ions (see PL04, and Liu, Petrosian \& Mason 2005a,
2005b). 

We have also evaluated the collisionless damping rate due to a
population of nonthermal electrons and protons. We have argued that
the most important process here is the {\it transit time damping} mechanism,
and show that this process gives a damping rate very similar to that
obtained for a thermal distribution. In general, the
damping rate is essentially proportional to the mean momentum  times
the number of the particles. Thus the relative importance of thermal
and nonthermal populations depends on the product of the ratios of their densities
and average momenta. In most cases except for extremely hard nonthermal tails
(electron index $a>-1.5$ or $-2$, for nonrelativistic and extreme
relativistic cases, respectively), this ratio will be less than the
energy content of the two population. In particular, this will be true
for most solar flares except for the strongest bursts where one
requires acceleration of all the available background electrons. This
behavior also indicates, that as  is the case of thermal damping, here
also the contribution of protons relative to electrons can be
neglected except for very rare flares with strong nuclear gamma-ray
line emission which require more energy for accelerated protons than
electrons. Most flares, however, are electron dominated and the
contribution of nonthermal electron will increase the damping rate by
the above basic ratio at all $k$ and $\theta$ and decrease the
value of the critical wavevector but not affect its anisotropy.

We stated that fast modes dominate slow and Alfvenic modes in terms of acceleration
and therefore concentrated on the fast modes. This is only true if 
turbulence is strong, 
i.e. when the critical balanced condition is satisfied for Alfv\'enic modes.
At the injection scale the turbulence may be weak (see Galtier et al. 2000)
and develop a cascade with $k_{\|}=const$. However, such a cascade has
a limited inertial range (see discussion in Lazarian \& Vishniac 1999,
Cho, Lazarian \& Vishniac 2003) and beyond this range it transfers to the strong
Alfv\'enic turbulence. Therefore disregarding weak turbulence seems
justified\footnote{More definite statements can be obtained if velocity 
fluctuations associated solar flares are analysed. The corresponding
techniques developed for the interstellar medium, e.g. Velocity Chanel
Analysis, Velocity Correlation Spectrum (Lazarian \& Pogosyan 2000, 2004),
Modified Velocity Centroids (Lazarian \& Esquivel 2003, Esquivel \& Lazarian
2005) can be applied for the purpose.}. Moreover, reconnection events should produce perturbations
with velocity of the order of Alfv\'en velocity, which should produce
strong turbulence from the very beginning.

MHD turbulence that we considered was balanced in the sense that the
equal flux of energy was assumed in every possible direction.
In solar corona we expect the energy injection to be localized both in
space and in time. As the result turbulent energy propagates from such
sources, e.g. reconnection region
creating an imbalanced cascade. The properties of imbalanced turbulence 
(see Maron \& Goldreich 2001, Cho, Lazarian
\& Vishniac 2002, Lithwick \& Goldreich 2003), in particular its damping 
time and scaling, can be very different from the balanced one.
The Alfv\'enic cascade is being strongly modified by imbalance and this
may result in much lower rates of cascading, if the imbalance is strong.
However, variations in Alfv\'en speed that are present in solar corona
are likely to result in reflecting Alfv\'en perturbations. These
reflections mitigate the imbalance and therefore we believe that the
effects of imbalance will not be substantial. A more detailed study of
the issue will be presented elsewhere. 

We have limited our considerations to scales above the particle
gyro radii, where the MHD approximation is valid. For shorter scales
one must consider mechanisms of damping or acceleration  other than the
TTD. In particular gyro-resonance scattering must be included with more
realistic dispersion relations than those given in equation
(\ref{disp}). We intend to address these extensions of the current results in our
future publications.

Transfer of turbulent energy from large to small scales and its damping is a general
process that can be important for heating and particle 
acceleration in various environments other than  solar flares, such as 
in gamma-ray bursts (see Lazarian et al. 2003) and
accretion around black holes (Liu, Petrosian \& Melia 2005).

We acknowledge support from the NSF grants ATM-0312282 at Wisconsin and  ATM-0312344 at Stanford.  
AL also acknowledge 
the NSF Center for Studies of Magnetic 
Self-Organization in Laboratory and Astrophysical Plasmas, AL and HY
acknowledge partial support from NSF AST-0307869,
and VP would like to acknowledge support from NASA grants NAG5-12111 and 
NAG5-11918-1. 

\appendix

\section{Damping Due To Ion Viscosity}

\label{viscousec}
Viscous damping is not important unless there is compression.
Therefore it only influences compressible modes and has marginal effects on Alfv\'{e}n modes. In a 
strong magnetic field, the proton gyrofrequency is much larger than proton collisional frequency 
$\tau_{\rm Coul}^{-1}\simeq c_S/\lambda_{\rm Coul}$, ($\Omega_p\tau_{\rm Coul}=5\times 
10^3(B/100{\rm G})(10^{10}{\rm cm}^{-3}/n)(T/10^7{\rm K})^{3/2})$, and 
the transport of transverse momentum is 
prohibited by the magnetic field. Thus, transverse
viscosity coefficient $\eta_{\perp}\sim\eta_{0}/(\Omega_{p}\tau_{\rm Coul})^{2}$ is much smaller than 
longitudinal viscosity coefficient $\eta_{0}=0.96nk_B T\tau_{\rm Coul}$ (see, {\it e.g.} 
Braginskii 1965).

Considering only the zeroth order terms due to longitudinal viscosity, the viscosity tensors are 
$\pi_{xx}=\pi_{yy}=-\eta_0(W_{xx}+W_{yy})/2$ and  $\pi_{zz}=-\eta_0 W_{zz}$, where $W_{jk}\equiv 
\frac{\partial v_j}{\partial x_k}+\frac{\partial v_k}{\partial x_j}-\frac{2}{3}\delta_{jk}\triangledown 
\centerdot {\bf v}$ is the rate-of-strain tensor, ${\bf v}$ is the fluid velocity (Sigmar 2002). 
Here, $z$ 
axis is defined by the magnetic field. Heat generated by the viscosity   is  

\be
Q_{vis}={\bf \pi}: \triangledown {\bf v}=-\pi_{xx}\frac{\partial v_x}{\partial x}-\pi_{xx}\frac{\partial v_y}{\partial y}-\pi_{zz}\frac{\partial v_z}{\partial z}=\eta_{0}(\partial v_{x}/\partial x+\partial v_{y}/\partial y-2\partial v_{z}/\partial z)^{2}/3.
\label{visheat}
\ee
Dividing this by the total energy associated with the fast modes, we obtain   the damping rate 
$\Gamma_{vis}$. While the damping due to compression along the magnetic fields (the 3rd term) 
can be easily understood, 
it is somewhat counterintuitive that the compression
perpendicular to magnetic field also results in damping through longitudinal
viscosity. However, the origin of this viscosity can be easily traced
(see Braginskii 1965). Indeed, for motions perpendicular to the magnetic
field $B$, $\triangledown_\perp \cdot{\bf v}=\dot{n}/n\sim\dot{B}/B$, 
implies the transverse energy of the ions increases due to the adiabatic 
invariant $v_{\perp}^{2}/B$. If the rate of compression
is faster than that of collisions, the ion distribution in the momentum
space will become distorted away from the isotropic Maxwellian sphere
to an oblate spheroid with the long axis perpendicular to the magnetic
field. As a result, the transverse pressure becomes greater than the
longitudinal pressure by a factor $\tau_{\rm Coul}\dot{n}/n$, resulting
in a stress $\sim P\tau_{\rm Coul}\dot{n}/n\sim\eta_{0}\triangledown_\perp \cdot v$, where $P=nk_BT$ is 
the longitudinal pressure.
The restoration of the equilibrium increases the entropy and causes
the dissipation of energy. In a low $\beta_p$ medium, compressions are perpendicular to magnetic field, 
thus $\Gamma_{vis}=k_\perp^2\eta_0/3nm_i$.
In high $\beta_p$ medium,  as
pointed out in $\S2$, the velocity perturbations are radial. Thus according to Eq.(\ref{visheat}), the 
corresponding damping rate $\Gamma_{vis}=k^{2}\eta_{0}(1-3\cos^{2}\theta)^{2}/(3nm_i)$.

Putting all these together for $m_i=m_p$ we get
\begin{eqnarray}
\Gamma_{vis}&=& 0.13(\beta_p^{1/2}/\tau_0)(kL)(k\lambda_{\rm Coul})\cases{\sin^2\theta,& 
for $\beta_p\ll 1,$\cr
(1-3\cos^{2}\theta)^{2},&  for $\beta_p\gg 1$\cr}
\label{viscous1}
\end{eqnarray}
or in terms of physical parameters appropriate for solar flares
\begin{eqnarray}
\Gamma_{vis}=0.05{\rm s}^{-1}(kL)^2\left(\frac{10^{8}{\rm cm}}{L}\right)^2\left(\frac{T}{10^7{\rm K}}\right)^{2.5}\left(\frac{10^{10}{\rm cm}^{-3}}{n}\right)\cases{\sin^2\theta& for $\beta_p\ll 1$\cr
(1-3\cos^{2}\theta)^{2}&  for $\beta_p\gg 1$\cr}
\label{viscous2}
\end{eqnarray}
This damping becomes important at a scale where $\Gamma_{vis}\tau_{cas}=1$. From the above
and equation (\ref{tcasfast}), we find
\be
k_cL=\left(\frac{L\beta_p}{18}\right)^{-\frac{1}{3}}\left(\frac{\lambda_{\rm Coul}\sin^2\theta}{M_A^2}\right)^{-\frac{2}{3}}=\frac{4}{\beta_p^{1/3}}\left(\frac{M_A}{\sin\theta}\right)^{4/3}\left(\frac{L}{10^8\rm cm}\right)^{2/3}\left(\frac{10^{10}\rm cm}{n}\right)^{-2/3}\left(\frac{10^7\rm K}{T}\right)
\ee

\section{General Nonthermal Damping Rates}

The nonthermal electron damping expression (\ref{erate1}) can be integrated for a power law 
distribution of electrons. The general expression is

\begin{eqnarray}
&&\Gamma_{nonth}({\bf k})=\frac{\pi}{8}\delta kc\left(\frac{1-\eta^2}{\eta}\right)\left(1-\frac{\beta_A^2}{\eta^2}\right)\frac{(a-1)N_0}{n} (\gamma_0-1)^{a-1}\times\nonumber\\
&&\cases{\frac{1}{32}\left\{\frac{(\gamma_m+1)^{1/2}}{\gamma_m-1}\left[32-\frac{\beta_A^2}{\eta^2}\left(26+\frac{8}{\gamma_m-1}\right)\right]+\left(16-19\frac{\beta_A^2}{\eta^2}\right) \sqrt{2}\,\tanh^{-1}\left(\sqrt{\frac{\gamma_m+1}{2}}\right)\right\} &
 for $a=2.5$,\cr
\frac{1}{15} \left\{(-5+7\,\frac{\beta_A^2}{\eta^2}\,)+\frac{\gamma_0\beta_0}{(\gamma_m-1)^3}\left[5(\gamma_0\beta_0)^2-\frac{\beta_A^2}{\eta^2}(7\,\gamma_m^2-6\gamma_m+2)\right]\right\} & for $a=3$,\cr
\frac{1}{384}\left\{\frac{(\gamma_m+1)^{1/2}}{\gamma_m-1}\left[48+\frac{192}{\gamma_m-1}-\frac{\beta_A^2}{\eta^2}\left(78+\frac{152}{\gamma_m-1}+\frac{64}{(\gamma_m-1)^2}\right)\right]\right.\cr
-\left.\left(24-39\frac{\beta_A^2}{\eta^2}\right) \sqrt{2}\,\tanh^{-1}\left(\sqrt{\frac{\gamma_m+1}{2}}\right)\right\} &
 for $a=3.5$,\cr
\frac{1}{105}\, \left\{(7-13\frac{\beta_A^2}{\eta^2}\,)-\frac{\gamma_0\beta_0}{(\gamma_m-1)^4}\left[7\,( \gamma_0\beta_0)^2(\gamma_m-4)
-\frac{\beta_A^2}{\eta^2}(13\gamma_m^3-52\gamma_m^2+32\,\gamma_m-8)\right]\right\} & for $a =4$,}
\end{eqnarray}
where $\gamma_i=1+E_i/(m_ec^2)$ are the Lorentz factors corresponding to kinetic energies $E_0$ 
and $E_m$.  In 
solar case, $\gamma_m-1\ll 1$, the above expression then simplifies to equation (\ref{eratenonrel}) in 
the main section. Similarly we can get
the extreme relativistic limit ($\gamma_m\gg1$) of equation (\ref{erateextrel}).

\end{document}